\documentclass[aps,prb,twocolumn,superscriptaddress,showpacs]{revtex4}
\usepackage{amsmath}
\usepackage{dcolumn}
\usepackage{amssymb}
\usepackage{amsfonts}
\usepackage{graphicx}
\usepackage[T1]{fontenc}
\usepackage{indentfirst}
\usepackage{color}

\begin{document}
\title{Fibonacci anyon excitations of one-dimensional dipolar lattice bosons}
\author{Tanja \DJ uri\'c}
\affiliation{Instytut Fizyki im. M. Smoluchowskiego, Uniwersytet Jagiello\'nski, \L{}ojasiewicza 11, 30-348 Krak\'ow, Poland}
\author{Krzysztof Biedro\'n}
\affiliation{Instytut Fizyki im. M. Smoluchowskiego, Uniwersytet Jagiello\'nski, \L{}ojasiewicza 11, 30-348 Krak\'ow, Poland}
\author{Jakub Zakrzewski}
\affiliation{Instytut Fizyki im. M. Smoluchowskiego, Uniwersytet Jagiello\'nski, \L{}ojasiewicza 11, 30-348 Krak\'ow, Poland}
\affiliation{Mark Kac Complex Systems Research Center, Jagiellonian University, \L{}ojasiewicza 11, 30-348 Krak\'ow, Poland}

\date{\today}
\begin{abstract}
We study a system of dipolar bosons in a one-dimensional optical lattice using exact diagonalization and density matrix renormalization group methods. In particular, we 
analyze low energy properties of the system at an average filling of 3/2 atoms per lattice site. We identify the region of the 
parameter space where the system has non-Abelian Fibonacci anyon excitations that correspond to fractional domain walls between different charge-density-waves. When such one-dimensional systems are 
combined into a two-dimensional network, braiding of Fibonacci anyon excitations has potential application for fault tolerant, universal, topological quantum computation.
Contrary to previous calculations, our results also demonstrate that super-solid phases are not present in the phase diagram for the discussed 3/2 average filling. Instead, decreasing the value of the nearest-neighbor tunneling strength leads to a direct, Berezinskii-Kosterlitz-Thouless, super-fluid to 
charge-density-wave quantum phase transition.
\end{abstract} 


\maketitle
\section{Introduction}
\label{sec:intro}
The large recent interest in non-Abelian topological phases of matter is strongly motivated by the possibility of a fault-tolerant topological quantum computation 
\cite{Kitaev,Freedman1,DasSarma,Nayak,Pachos,Pachos2}
based upon non-Abelian anyons \cite{Moore,Stern1,Stern2,Trebst}  that appear as quasiparticle excitations for such exotic quantum phases of matter. The errors caused by 
local interactions with the environment are a basic obstacle for quantum computation. 
The main idea behind topological quantum computation is that non-Abelian anyonic quasiparticles can be used to encode and manipulate information in a way that
is resistant to errors, and therefore to perform fault-tolerant quantum computation without loss of information (decoherence).

The understanding of the origin and properties of  non-Abelian states of matter is also of fundamental importance and is at the frontier of current theoretical 
and experimental research \cite{DasSarma2,Stern3,Bonderson,Zhu1,Zhu2}.  The main objective is the investigation of new models that have non-Abelian quasiparticle excitations, or support non-Abelian defects,  
as a result of complex interplay between topology and quantum mechanics \cite{Kitaev2,Teo}. The robustness against small local perturbations is due to the topological nature of these states 
of matter, that therefore can be used as building blocks for topological quantum computation.

In this paper we study a system of ultra-cold dipolar bosons trapped in a one-dimensional (1D) optical lattice and at an average filling of $3/2$ atoms per lattice site. The system can be well described by an
extended Bose-Hubbard Hamiltonian with the on-site and nearest-neighbor interactions \cite{Wikberg}. We study the ground states and low energy elementary excitations of the system 
in the regime of small tunneling between lattice sites and identify the region of the parameter space where the system supports non-Abelian, SU(2)$_3$ Fibonacci anyon excitations.

In 1D quantum statistics is not well defined. The interchange of two quasiparticles in one spatial dimension is impossible without one particle going through another. Therefore the 
adiabatic exchange (braiding) of these quasiparticles is not possible in the strictly 1D system that we have considered. However, braiding can be achieved by connecting 
these 1D systems with T-junctions into a two-dimensional (2D) network as suggested previously in the case of Majorana quantum wires \cite{Alicea}.

Our results show that the system supports Fibonacci anyon excitations in the regime where (quasi)degenerate manifolds of energy states are well defined, without crossings 
between the energy levels within different manifolds. This regime corresponds only to a part of the charge-density-wave (CDW) region in the phase diagram of the system, while the system 
has nontrivially (quasi)degenerate ground states in the whole CDW region. As indicated in previous studies \cite{Wikberg,Ardonne}, Fibonacci anyon excitations correspond to 
fractional domain walls between different CDWs.

Also, contrary to previous calculations based on Gutzwiller wavefunction approach \cite{Wikberg}, our results demonstrate that the 
super-solid (SS) phases are not present between the super-fluid (SF) and CDW regions of the phase diagram of the system for the specific average filling of 3/2 atoms per lattice site considered throughout this paper. 
The system for arbitrary fillings has also been considered by Batrouni \emph{et al.} \cite{Batrouni} where the SS phases were observed at other higher fillings. Still at the particular 
value of $\nu=3/2$ the autors of \cite{Batrouni} were unable to verify the presence of the SS phases. We claim that instead, decreasing the tunneling strength between the neighboring sites leads to a direct, Berezinskii-Kosterlitz-Thouless (BKT), super-fluid (SF) to 
charge-density-wave (CDW) quantum phase transition.

The anyonic quasiparticles, which are neither fermions nor bosons, are associated to systems in two spacial dimensions. Namely, when two quasiparticles are exchanged in 
two dimensions, the wavefunction of the system can gain any phase factor $e^{i\alpha}$, which motivated the name  anyons. On the other hand in three spacial dimensions the only possible phase factors are 
$e^{i\alpha}=+1$ or $-1$, which corresponds to bosons or fermions.

If in addition there are $m$ degenerate states $\psi_i$ ($i=1$,...,$m$) for $n$ quasiparticles at positions 
$x_1$,...,$x_n$, the result of the quasiparticle exchanges is more than just a change of the phase of the wavefunction. In that case an exchange of two quasiparticles can rotate one 
of the degenerate states, $\psi_i$, into a different degenerate state $\psi_j$ within a $m$-dimensional 
degenerate Hilbert space for $n$ quasiparticles, $\psi_i\rightarrow A_{ij}\psi_j$. In general, exchange of other two quasiparticles will be described by a different rotation matrix,
$\psi_i\rightarrow B_{ij}\psi_j$. For two consecutive exchanges of the quasiparticles, the final state of the system will depend upon the order in which these exchanges 
were performed, since the matrices $A$ and $B$ do not commute, that is $AB\neq BA$. Such states and their quasiparticle excitations are therefore called non-Abelian or non-commutative.

This exotic non-Abelian statistical behaviour allows fault-tolerant manipulation of the quantum information stored in $m$-dimensional Hilbert space of $n$ non-Abelian quasiparticles. 
Quantum computation is a process of initializing a controllable quantum system to some known initial state $|\psi_i\rangle$, evolving system by a unitary transformation 
$U(t)$ to some final state $|\psi_f\rangle$, and finally measuring the state $|\psi_f\rangle$ at the end of the computation. The quantum computational code is defined by 
the unitary transformations, which can be engineered to be any unitary transformations if there is sufficient control over the underlying Hamiltonian of the system.

For a large 
class of non-Abelian states any unitary transformation can be generated only by braiding quasiparticles \cite{Freedman1,DasSarma}, which consequently allows universal topological quantum computation 
through braiding. An example of such non-Abelian states are the states that support SU(2)$_3$ Fibonacci anyon quasiparticle excitations \cite{Freedman1,Nayak,DasSarma}. The final result of the 
computation, that is the final state of the system after evolution by a unitary transformation, can be obtained by a topological measurement based on a non-Abelian generalization 
of the Aharonov-Bohm effect \cite{Freedman1,Nayak,DasSarma}.

Non-Abelian states were initially predicted in fractional quantum Hall (FQH) systems \cite{Moore,Greiter2,Nayak2,ReadRezayi1,ReadRezayi2,Wilkin,Cappelli} that are constrained to two spacial dimensions, and subsequently in various similar 
FQH-like systems in 2D \cite{Nayak,Wilkin,Cooper,Gurarie,Moller,Duric3,Wu,Sterdyniak,Fu1,Fu2,Nilsson,Nikolic,Sau,Levin}. However, analogous states were also found to appear in various one-dimensional (1D) models \cite{Tsvelik1, 
Tsvelik2,Fendley,Tu,Paredes1,Paredes2,Nielsen,Greiter,Thomale,Duric,Alicea2}.

Whether in 1D or 2D, non-Abelian states of matter have a global hidden order with constituent particles following a global pattern that is not associated with breaking of any 
symmetry. This hidden order is associated with organization of particles in indistinguishable clusters \cite{Cappelli,Paredes1,Paredes2,Duric,Paredes3}. Each cluster corresponds to an underlying Abelian copy, 
and SU(2)$_k$ non-Abelian states can be obtained from $k$ such Abelian copies by symmetrizing over the 
coordinates of the clusters \cite{Cappelli,Paredes1,Paredes2,Duric,Paredes3}. This symmetrization (indistinguishability) can be achieved by applying a projection operator to a direct product of the wavefunctions 
for $k$ copies, which introduces the possibility of topological degeneracy and non-Abelian statistics in the space of quasiparticles.

The projection operator projects $k$ local degrees of freedom corresponding to $k$ copies onto a new degree of freedom that is symmetric under exchange of any of the 
$k$ components, and leads to a topological degeneracy not related to simple symmetry considerations. This topological degeneracy is robust against perturbations and interactions 
with the environment.

In our calculations we use exact diagonalization (ED) and density matrix renormalization group (DMRG) \cite{White,Schollwock,ITensor} methods to study low energy properties of the system for system 
sizes up to 50 lattice sites and with periodic boundary conditions.

The region of the parameter space where the system supports non-Abelian Fibonacci anyon excitations is determined by calculating the overlaps between the exact wave functions 
for the low-energy states of the Hamiltonian describing the system at average filling of $\nu=3/2$ atoms per lattice site, and the corresponding ansatz wavefunctions which 
have SU(2)$_3$ non-Abelian order by construction.

The ansatz states are constructed by applying a symmetrization projection operator to a direct product of the corresponding 
wave functions for three Abelian copies at filling fraction $\nu=1/2$ atoms per lattice site. The projection operator introduces indistinguishability between the copies 
(symmetrization over the coordinates of the clusters) which leads to SU(2)$_3$ non-Abelian order.

The paper is organized as follows. In Sec. \ref{sec:Fibonacci_anyons} we consider exactly solvable points in the parameter space of the underlying extended Bose-Hubbard Hamiltonian, and demonstrate 
that Fibonacci anyon excitations correspond to fractional domain walls between different degenerate CDW ground states of the system. In Sec. \ref{sec:ED_DMRG} we present ED and DMRG results 
away from the exactly solvable points. In Sec. \ref{sec:SF_CDW}  we further characterize the SF to CDW quantum phase transition.
Protocol for braiding fractional domain walls within a 2D T-junction network is described in Sec. \ref{sec:Braiding}. 
We draw our conclusions in the final section, Sec. \ref{sec:Conculsions}. 

\section{Fibonacci anyon excitations as fractional domain walls}
\label{sec:Fibonacci_anyons}
The system of ultra-cold dipolar bosons in a 1D optical lattice can be well described by an extended Bose-Hubbard Hamiltonian of the form \cite{Wikberg}
\begin{eqnarray} \label{eq:H_EBH}
 H &=& -t\sum_i\left(a_i^{\dagger}a_{i+1}+a_{i+1}^\dagger a_i\right)+\frac{U}{2}\sum_i n_i(n_i-1) \nonumber \\
 &+&V\sum_i n_i n_{i+1},
\end{eqnarray}
where $t$ is the tunneling amplitude between the neighboring sites, $U$ is the on-site interaction, $V$ is the nearest-neighbor interaction, and the bosonic operators 
$a_i^\dagger$/$a_i$ create/annihilate a boson on site i. The operator $n_i=a_i^\dagger a_i$  denotes the number of bosons on site $i$.

Previous studies showed that the Hamiltonian (\ref{eq:H_EBH}) near the lattice filling $\nu=k/2$ supports SU(2)$_k$ anyonic excitations in the parameter regions where 
the system has nontrivially degenerate CDW ground states \cite{Wikberg, Ardonne}. In particular, at average filling $\nu=3/2$ that we consider, the low energy 
excitations are SU(2)$_3$ Fibonacci anyons. To demonstrate that the low-energy SU(2)$_k$ anyonic excitations correspond to domain walls between different degenerate CDW ground states we first 
consider exactly solvable points in the parameter space, that is, the ground states and the low energy excitations of the Hamiltonian (\ref{eq:H_EBH}) at 
$t=0$ and $U=2V$.

In general, for the filling fraction $\nu = k/2$ at $t=0$ and $U=2V$, the ground state has nontrivial degeneracy. The ground states are all CDW states with unit cells $[l,k-l]$, where $l=0,1,...,k$ \cite{Wikberg, Ardonne}.
For $\nu=3/2$ there are four degenerate CDW ground states:
\begin{eqnarray}\label{eq:CDWs}
|030303...\rangle&\equiv&[03] \\
|121212...\rangle&\equiv&[12] \nonumber \\
|212121...\rangle&\equiv&[21] \nonumber \\
|303030...\rangle&\equiv&[30]. \nonumber 
\end{eqnarray}
The low energy quasiparticle/quasihole excitations correspond to domain walls between degenerate CDWs with unit cells $[l,k-l]$ and $[l\pm1,k-l\mp1]$ \cite{Wikberg, Ardonne}.
More precisely, following domain walls correspond to elementary excitations \cite{Ardonne}
\begin{eqnarray}\label{eq:Domain_walls}
&&[k-l,l][k-l-1,l+1]\mbox{\;\;for\;\;} 0<l<k, \\
&&[k-l,l][k-l+1,l-1]\mbox{\;\;for\;\;} 0<l<k,  \nonumber \\
&&[k,0][k-1,1], \nonumber \\
&&[0,k][1,k-1], \nonumber
\end{eqnarray}
where $[a,b][c,d]\equiv |...ababcdcd ...\rangle$. For the filling fraction $\nu=3/2$ elementary quasihole and quasiparticle excitations are \cite{Ardonne} 
\begin{eqnarray}\label{eq:Fibonacci_domain_walls}
|...21211212...\rangle&\equiv&[21][12], \\
|...12122121...\rangle&\equiv&[12][21],\nonumber \\
|...12120303...\rangle&\equiv&[12][03],\nonumber\\
|...21213030...\rangle&\equiv&[21][30].\nonumber
\end{eqnarray}
If the lattice bosons have a charge $q$, than the quasiparticle/quasihole excitations have a fractional charge $\pm q/2$ \cite{Wikberg, Ardonne}. In other words, the states (\ref{eq:Fibonacci_domain_walls}) have one boson more or less
at two sites where a domain wall is formed. Since the states ($\ref{eq:Fibonacci_domain_walls}$) have one particle more or less than the ground states, for a system with fixed number 
of particles the elementary excitations are quasiparticle-quasihole pairs.

We further demonstrate that these fractional domain walls are non-Abelian SU(2)$_3$ Fibonacci anyons \cite{Trebst}, similar to the elementary excitations of the $\nu=12/5$ 
Read-Rezayi fractional quantum Hall (FQH) state \cite{ReadRezayi1,ReadRezayi2}. If a fractional domain wall is a Fibonacci anyon then its quantum dimension is the golden ratio $d_F=(1+\sqrt{5})/2$ \cite{Trebst,Vaezi1, Vaezi2}. The Fibonacci 
sequence is a sequence with the property that each number in the sequence is the sum of the previous two numbers in the sequence. The non-Abelian anyons with quantum dimension equal to golden ratio are named Fibonacci anyons 
because the ratio of any number in the Fibonacci sequence to the previous number in the sequence is approximately the golden ratio.

The quantum dimension for these fractional domain walls can be found by considering an adjacency matrix for the elementary excitations \cite{Vaezi1}. We first note that here charge $q/2$ and charge $-q/2$ 
elementary excitations are topologically equivalent excitations because they differ by a local operator \cite{Vaezi1}. The adjacency matrix can then be obtained by considering 
which pairs of ground states create a $\pm q/2$ fractional domain wall and is given by 
\begin{equation}\label{eq:adjacency_matrix}
 A = \left(\begin{array}{cccc}
0 & 1 & 1 & 0 \\
1 & 0 & 0 & 1 \\
1 & 0 & 0 & 0 \\
0 & 1 & 0 & 0 \end{array} \right),
 \end{equation}
where the rows/columns $1,2,3$ and $4$ refer to the $[21]$,$[12]$,$[30]$ and $[03]$ ground states, respectively.

The adjacency matrix (\ref{eq:adjacency_matrix}) encodes fusion rules for the elementary excitations \cite{Vaezi1,Vaezi2}
\begin{equation}\label{eq:Fusion_rules}
i\times j = \sum_k (A_i)_{jk} k, 
\end{equation}
where $A_i$ is the adjacency matrix of the quasiparticle $i$. These fusion rules determine the number of ways that quasiparticles $i$ and $j$ can fuse into quasiparticle $k$. 
For the Fibonacci anyons $\tau$ the fusion rule is 
\begin{equation}\label{eq:Fusion_rules_Fibonacci_anyons}
\tau\times\tau = 1 +\tau. 
\end{equation}

Due to the Fibonacci anyon algebra (\ref{eq:Fusion_rules_Fibonacci_anyons}) the ground-state degeneracy in the presence of $n$ Fibonacci anyon excitations satisfies the 
Fibonacci recursion relation \cite{Vaezi2}
\begin{equation}\label{Fibonacci_recursion_relation}
G(n)=G(n-1)+G(n-2). 
\end{equation}
In the large $n$ limit the ground-state degeneracy grows as \cite{Vaezi1, Vaezi2}
\begin{equation}
\log G(n) \sim n\log d_F ...  
\end{equation}
with $d_F$ being the Fibonacci anyon quantum dimension that corresponds to the maximum eigenvalue of the adjacency matrix (\ref{eq:adjacency_matrix}) \cite{Vaezi1}.

For a $m$-fold degenerate ground-state manifold the statistics of anyons can be described by $m\times m$ unitary matrices that act on the ground-state manifold. 
Since $m\times m$ unitary matrices form a non-Abelian group (matrices $A$ and $B$ generally do not commute, $ AB \neq BA$), these anyons are called non-Abelian anyons.

In the parameter region where the system supports non-Abelian elementary excitations the ground-state degeneracy depends on the topology of the manifold on which the 
system is defined. For the lattice filling $\nu = k/2$ the ground-state will be $k+1$-fold degenerate for periodic boundary condition and non-degenerate 
for open boundary condition. In other words, the system has non-trivial non-Abelian topological order reflected in topological ground-state degeneracy \cite{Oshikawa1, Oshikawa2}. We also 
note that in general a topologically ordered state has a quasi-degenerate ground state manifold for a finite system size that becomes exactly degenerate in the 
thermodynamic limit. That will be the case away from the exactly solvable points ($U=2V$, $t=0$) as described in the following section. 
\section{Numerical results} 
\label{sec:ED_DMRG}
\begin{figure}[t!]
 \includegraphics[width=\columnwidth]{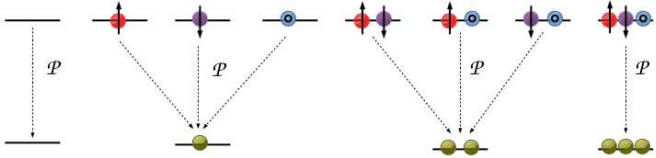}
\caption{\label{fig:P_operator} (Color online) Schematic of the local projection operator $\mathcal{P}_i$ at
a lattice site $i$. The operator $\mathcal{P}_i$ projects the three local degrees of freedom $\uparrow$, $\downarrow$ and $\circ$, onto a new degree of freedom that is
symmetric under exchange of any of the three components. In other words,  $\mathcal{P}_i$ maps the single site $8$-dimensional Hilbert space of
three species of hard-core bosons $\uparrow$ (red spheres), $\downarrow$ (purple spheres) and $\circ$ (blue spheres) to the
single-site $4$-dimensional Hilbert space of four-hard-core-bosons (green spheres). These four-hard-core bosons obey generalized exclusion principle - less than four bosons at any lattice site $i$. 
  } 
 \end{figure}
To study properties of the system away from the exactly solvable points ($U=2V$, $t=0$) we use ED and DMRG \cite{White,Schollwock, ITensor} methods. Validity of our DMRG results 
is confirmed by comparison with the ED results for smaller system sizes ($L\leq 14$ lattice sites).

We primarily study the ground states and low-lying excitations of the system with periodic boundary conditions for $U,V \gg t$ and for a fixed number of atoms, 
$N=3L/2$. For such states large occupation of a single site is improbable. This allows 
the local Hilbert space truncation to single site Fock states $|n_i\rangle$ containing at most $n=n_{max}$ atoms. For the lattice filling $\nu=N/L=3/2$ it is sufficient 
to take $n_{max}=3$, that is, the local Hilbert space of dimension four with $n_i=0,1,2,3$.

We first demonstrate that there is a parameter region where the system supports non-Abelian excitations. For those parameter values all 
(quasi)degenerate lowest energy states have a high overlap ($\simeq 1$) 
with the corresponding manifold of four ansatz states that have SU(2)$_3$ non-Abelian topological order by construction.
We also show that elementary excitations above such states exhibit non-Abelian statistics. 

\begin{figure}[b!]
 \includegraphics[width=\columnwidth]{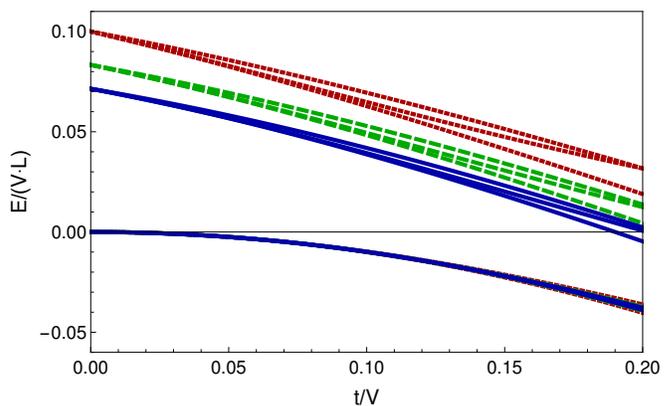}
\caption{\label{fig:E12_1} (Color online) The ED results for the first five energy levels of the Hamiltonian (\ref{eq:H_sigma}) at filling fraction $\nu=1/2$ and 
with periodic boundary conditions, as functions of the tunneling parameter $t/V$ (with $V$ being the nearest-neighbor interaction) and for the 
 system sizes $L=10$ (red dotted lines), 12 (green dashed lines) and 14 (blue solid lines) lattice sites. Here the energy values (per lattice site) are in units of V.
  } 
 \end{figure}

The four non-Abelian ansatz states for the lowest energy, (quasi)degenerate manifold at filling fraction $\nu=3/2$ can be constructed from the two
lowest energy, (quasi)degenerate, Abelian states at filling fraction $\nu=1/2$, $|\phi_{\sigma}^{(k)}\rangle_{\bar{t}}$ ($k=1,2$), by orthonormalization of the following 
wave-functions subspace \cite{Paredes1,Paredes2,Paredes3,Duric}:
\begin{equation}\label{eq:Ansatz1}
|\psi^{(l,m,n)}\rangle_{(\bar{t},\bar{U})} = \mathcal{P} \left(|\phi_{\uparrow}^{(l)}\rangle_{\bar{t}}\otimes|\phi_{\downarrow}^{(m)}\rangle_{\bar{t}}\otimes|\phi_{\circ}^{(n)}\rangle_{\bar{t}}\right), 
\end{equation}
where $l,m,n=1,2$ and $\sigma = \uparrow,\downarrow,\circ $ denotes three $\nu=1/2$ copies. The tunneling parameter and the on-site interaction strength are 
denoted by $\bar{t}=t/V$ and $\bar{U}=U/V$, respectively.

Here the wave-functions $|\phi_{\sigma}^{(k)}\rangle_{\bar{t}}$ ($k=1,2$) correspond to the 
two lowest energy (quasi)degenerate states of the Hamiltonian 
\begin{eqnarray}\label{eq:H_sigma}
H_{\sigma}&=& -t\sum_{i} \left(a_{\sigma,i}^{\dagger}a_{\sigma,i+1}+a_{\sigma,i+1}^{\dagger}a_{i,\sigma}\right) \nonumber \\
&+&V\sum_{i} n_{\sigma,i}n_{\sigma,i+1},
\end{eqnarray}
at average filling $\nu=1/2$ atoms per lattice site and with periodic boundary conditions, $n_{\sigma,i} = a^{\dagger}_{\sigma,i}a_{\sigma,i}$ and $a_{\sigma,i}^{\dagger}$/$a_{\sigma,i}$ 
are hard-core boson creation/annihilation operators at site $i$
satisfying $(a_{\sigma,i}^{\dagger})^2=0$ (that is, only allowed occupation numbers are $n_i^{\sigma}=0$ or $1$ bosons per site). 
 \begin{figure}[t!]
 \includegraphics[width=\columnwidth]{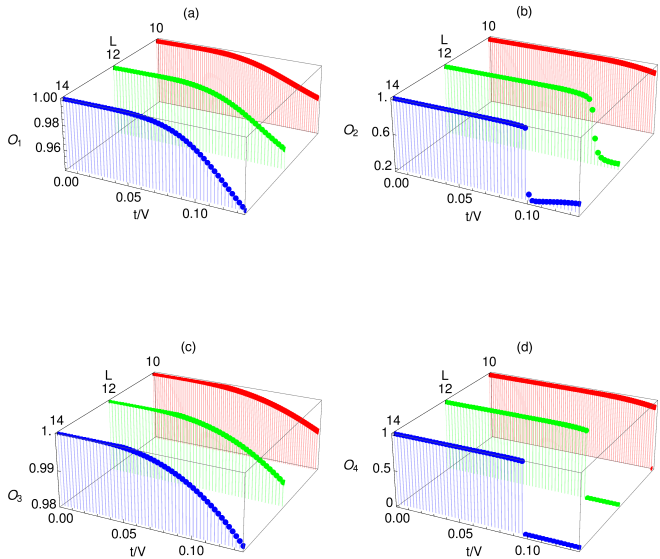}
\caption{\label{fig:Overlap_U2}(Color online) 
The ED results for the total overlaps (\ref{eq:Overlap1}) of the four 
lowest energy, (quasi)degenerate, exact ground states of the Hamiltonian (\ref{eq:H_EBH}) at average filling of $\nu=3/2$ atoms per lattice site and with periodic boundary conditions 
((a)-(d)), with the corresponding orthonormalized
ansatz states. Here $t/V$ is the tunneling parameter with $V$ being the nearest-neighbor interaction, and the on-site 
interaction strength is $U/V=2$. The system sizes are $L=10$, 12 and 14 sites (red, green and blue symbols, respectively).} 
 \end{figure}
 \begin{figure}[b!]
  \includegraphics[width=\columnwidth]{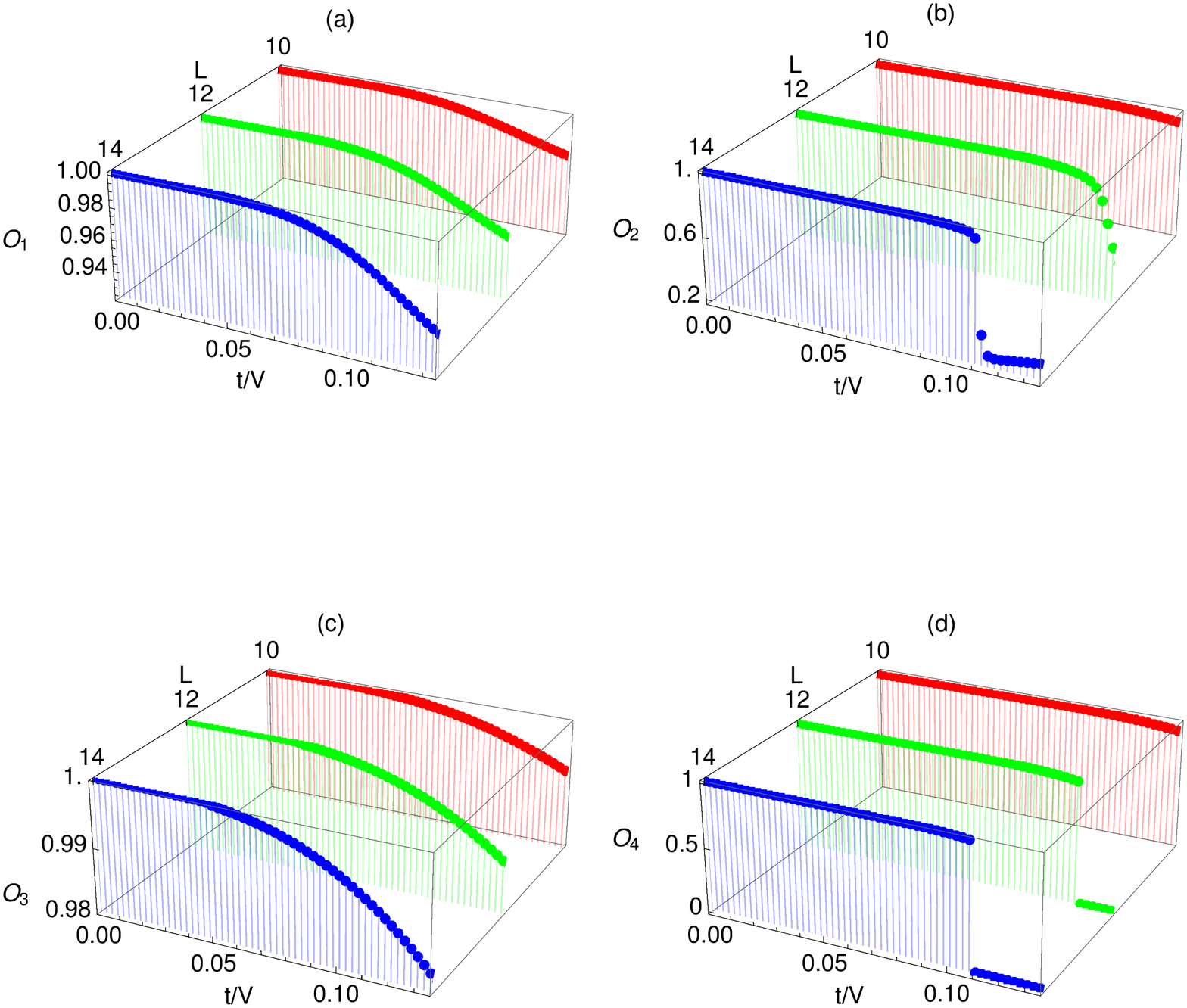}
 \caption{\label{fig:Overlap_U199}Same as Fig. \ref{fig:Overlap_U2} for the on-site interaction strength $U/V=1.99$.} 
  \end{figure}
 \begin{figure}[t!]
  \includegraphics[width=\columnwidth]{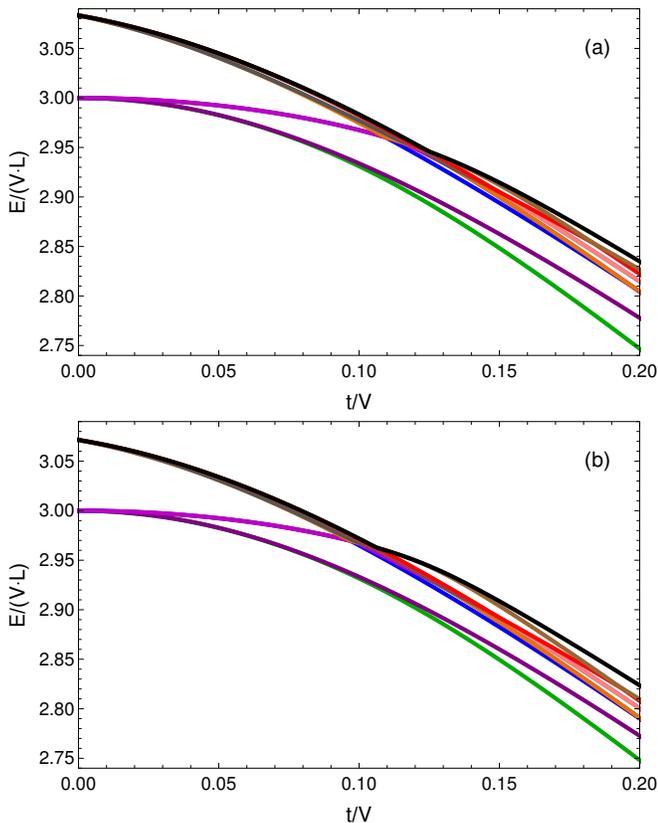}
 \caption{\label{fig:EnergyLevels_U=2V}(Color online)
 The ED results for the first ten energy levels of the Hamiltonian (\ref{eq:H_EBH}) at filling fraction $\nu=3/2$ and with periodic boundary conditions, as functions of the 
 tunneling parameter $t/V$ and for the system sizes (a) L=12 and (b) L=14 lattice sites. Here the on-site interaction strength is $U/V=2$, with $V$ being the nearest-neighbor 
 interaction. 
   } 
 \end{figure}

At $t=0$ the wave-functions $|\phi_{\sigma}^{(k)}\rangle_{\bar{t}=0}$ ($k=1,2$) are two degenerate CDW states with unit cells $[0,1]$ and $[1,0]$ and the low energy excitations of the Hamiltonian 
(\ref{eq:H_sigma}) are $\pm q/2$ fractional domain walls that are Abelian anyons similar to the quasiparticle and quasihole excitations of the $\nu=1/2$ Laughlin FQH state \cite{Laughlin}. As illustrated in Fig. \ref{fig:E12_1}, 
the states $|\phi_{\sigma}^{(k)}\rangle_{\bar{t}}$ ($k=1,2$)
at some finite value of the parameter $\bar{t}=t/V$ are adiabatically connected to the states at $t=0$, and therefore have Abelian topological order.

The projection operator $\mathcal{P}$ has the form 
\begin{equation}\label{eq:P}
\mathcal{P}=\mathcal{P}_i^{\otimes L},
\end{equation}
with $L$ being the number of lattice sites. Here $\mathcal{P}_i$ is the local projection operator at a lattice site $i$,
\begin{equation}\label{eq:Pi}
 \mathcal{P}_i= \left(\begin{array}{cccccccc}
1 & 0 & 0 & 0 & 0 & 0 & 0 & 0\\
0 & 1 & 1 & 1 & 0 & 0 & 0 & 0\\
0 & 0 & 0 & 0 & \sqrt{2} & \sqrt{2} & \sqrt{2} & 0\\
0 & 0 & 0 & 0 & 0 & 0 & 0 & \sqrt{6} \end{array} \right),
 \end{equation}
$\mathcal{P}_i$ maps $8$-dimensional Hilbert space of three species of
hard-core bosons, $\uparrow$, $\downarrow$ and $\circ$, to the single-site $4$-dimensional Hilbert space of four-hardcore bosons that obey generalized exclusion principle -
less than four bosons at any site $i$, as illustrated in Fig. \ref{fig:P_operator}.

After orthonormalization of the wave-functions subspace (\ref{eq:Ansatz1}) we find four linearly independent ansatz states, denoted here by 
$|\psi_{Ansatz}^{(k)}\rangle_{(\bar{t},\bar{U})}$. The number of linearly independent ansatz states corresponds to the 
number of lowest energy, (quasi)degenerate states of the Hamiltonian (\ref{eq:H_EBH}) that form the ground state manifold of the Hamiltonian (\ref{eq:H_EBH}).

The states $|\psi_{Ansatz}^{(k)}\rangle_{(\bar{t},\bar{U})}$, ($k=1,2,3,4$) form an orthonormal basis within (quasi)degenerate manifold, which leads to 
the following expression for the total overlap with the exact lowest energy (quasi)degenerate states of the Hamiltonian (\ref{eq:H_EBH}):
\begin{equation}\label{eq:Overlap1}
O_{i,(\bar{t},\bar{U})}= \sqrt{\sum_{k=1}^4 |_{(\bar{t},\bar{U})}\langle\psi_{Exact}^{(i)}|\psi_{Ansatz}^{(k)}\rangle_{(\bar{t},\bar{U})}|^2},
\end{equation}
where $i=1,...,4$. The ED results for the overlaps (\ref{eq:Overlap1}) for the system sizes $L=10,12$ and $14$ lattice sites are shown in Fig. \ref{fig:Overlap_U2} 
and Fig. \ref{fig:Overlap_U199}. The figures show overlaps for the four lowest (quasi)degenerate states (ground state manifold) of the Hamiltonian (\ref{eq:H_EBH}) for a range of values of the 
tunneling parameter $t/V$ and for two values of the on-site interaction strength, $U/V=2$ and $U/V=1.99$.

For $U=2V$ and $t=0$ (exactly solvable points) these states are four degenerate CDWs with unit cells $[03]$, $[30]$, $[12]$ and $[21]$, and the overlaps are exactly $1$. This reflects
non-Abelian nature of these states since the ansatz wave-functions have non-Abelian topological order by construction, and is in agreement with the results discussed in the previous section.
However, the overlaps for all four states are $\simeq 1$ for a range of values of the tunneling parameter $t/V$, both at $U=2V$ (Fig. \ref{fig:Overlap_U2}) and slightly away from $U=2V$ (for example for $U=1.99V$, Fig.  \ref{fig:Overlap_U199}).
This indicates non-Abelian nature of the states away from the exactly solvable points.

\begin{figure}[b!]
 \includegraphics[width=\columnwidth]{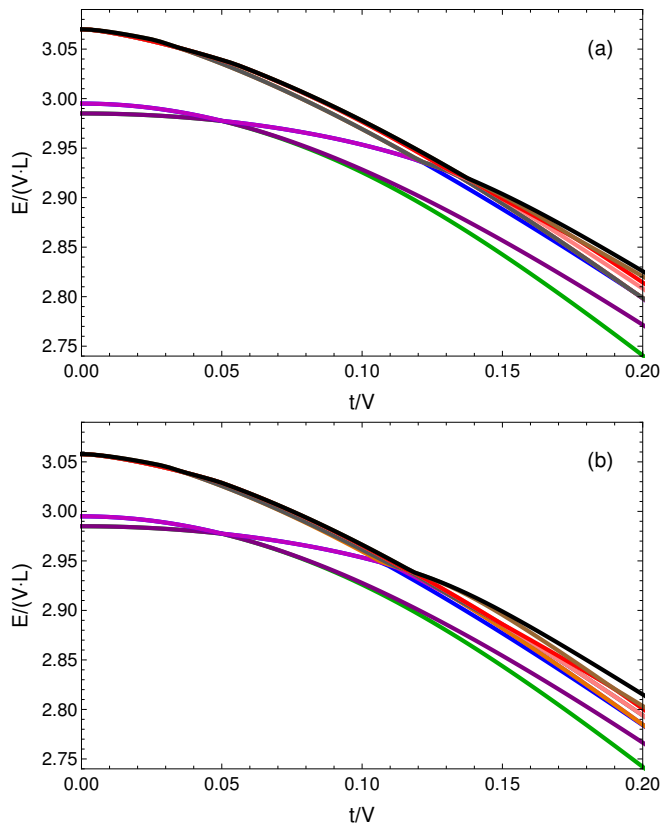}
\caption{\label{fig:EnergyLevels_U=1.99V}Same as Fig. \ref{fig:EnergyLevels_U=2V} for the on-site interaction strength $U/V=1.99$.
  } 
 \end{figure}

\begin{figure}[t!]
 \includegraphics[width=\columnwidth]{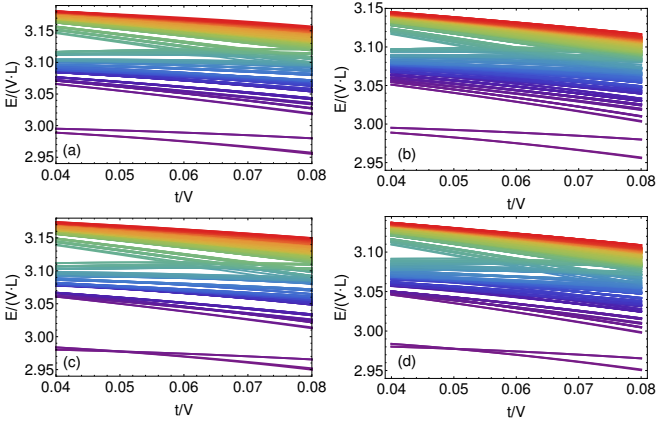}
\caption{\label{fig:EnergyLevels_L10_L12} (Color online) The energy levels of the Hamiltonian (\ref{eq:H_EBH}) at average filling of $\nu=3/2$ atoms per lattice site, 
obtained by ED method for the system sizes $L=10$ ((a) and (c)) and $L=12$ ((b) and (d)) lattice sites and with periodic boundary conditions. Here on-site interaction 
strength $\bar{U}=U/V=2$ ((a) and (b)) and $\bar{U}=U/V=1.99$ ((c) and (d)), with $V$ being the nearest-neighbor interaction. As explained in the text the system supports 
Fibonacci anyon excitations in the regime $(t/V) \lesssim 0.05$ where (quasi)degenerate energy manifolds are well defined and there is no level crossing between the states within  
different manifolds. 
 } 
 \end{figure}

Sudden decrease of the overlap, from $\simeq 1$ to zero, for the states (b) and (d) in Fig. \ref{fig:Overlap_U2} and Fig. \ref{fig:Overlap_U199}, is related to a crossing between the energy levels 
within the (quasi)degenerate, ground state manifold, and the energy levels within the (quasi)degenerate first excited manifold. That can be clearly seen in Fig. \ref{fig:EnergyLevels_U=2V} and Fig. \ref{fig:EnergyLevels_U=1.99V}.  
For the states (a) and (c) in Fig. \ref{fig:Overlap_U2} and Fig. \ref{fig:Overlap_U199} the overlaps start deceasing away from $\simeq 1$ at some value of $t/V=\bar{t}_c(L)$. The value 
$\bar{t}_c$ is characterized by a crossing between the energy levels within the (quasi)degenerate, first excited states manifold, and the energy levels within the (quasi)degenerate, 
second excited states manifold. These level crossings for the system sizes $L=10$ and 12 are shown in Fig. \ref{fig:EnergyLevels_L10_L12}.

To confirm non-Abelian nature of the states for $(t/V)<\bar{t}_c$, we further study elementary excitations above the (quasi)degenerate ground state manifold. By construction, the ansatz 
states (\ref{eq:Ansatz1}) have a hidden global order associated with the organization of the particles in three copies of $\nu=1/2$ states ($\uparrow$,$\downarrow$,$\circ$). The elementary excitations 
can be constructed by considering the elementary excitations of the three $\nu=1/2$ copies and symmetrizing \cite{Paredes3,Duric}. Non-Abelian statistics appears 
as a consequence of the symmetrization (introduced with projection operator $\mathcal{P}$)  which leads to a topological degeneracy in the subspace of elementary excitations and non-Abelian algebra of exchanges of elementary 
excitations (domain walls) \cite{Paredes3}.

The ansatz states for the first excited states manifold can be constructed by orthonormalization of the following wave-functions subspace \cite{Paredes3,Duric}
\begin{equation}\label{eq:Ansatz2}
|\bar{\psi}^{(l,m,n)}\rangle_{(\bar{t},\bar{U})} = \mathcal{P} \left(|\phi_{\uparrow}^{(l)}\rangle_{\bar{t}}\otimes|\phi_{\downarrow}^{(m)}\rangle_{\bar{t}}\otimes|\bar{\phi}_{\circ}^{(n)}\rangle_{\bar{t}}\right), 
\end{equation}
where $l,m=1,2$ and $n=L(L/2-1)$ with $L$ being the number of lattice sites. Here the wave functions $|\phi_{\sigma}^{(k)}\rangle_{\bar{t}}$ ($k=1,2$) correspond to 
the two lowest energy (quasi)degenerate states of the Hamiltonian (\ref{eq:H_sigma}) at average filling $\nu=1/2$, and the wave functions $|\bar{\phi}_{\sigma}^{(n)}\rangle_{\bar{t}}$ 
correspond to the states within the (quasi)degenerate, first excited states manifold of the Hamiltonian (\ref{eq:H_sigma}) at $\nu=1/2$.

The elementary excitations of the Hamiltonian (\ref{eq:H_sigma}) at $\nu=1/2$ and for a fixed number of particles are $\pm q/2$ domain wall pairs (quasiparticle-quasihole pairs) 
of the type $[01][10]$-$[10][01]$. The number of states in the first excited manifold at $\nu=1/2$, $\bar{N}=L(L/2-1)$), corresponds to the number of different pairs of sites $(i,j)$ 
where the domain walls can be created. In addition, there are three possible choices of the two ground states in the ansatz (\ref{eq:Ansatz2}): ($l=1$,$m=1$) , ($l=1$,$m=2$) and ($l=2$,$m=2$), which gives 
in total $\bar{N}_L=3L(L/2-1)$ linearly independent ansatz states for the first excited states manifold at $\nu=3/2$. These ansatz states, 
denoted by $|\bar{\psi}_{Ansatz}^{(k)}\rangle_{(\bar{t},\bar{U})}$ ($k=1$,2,...,$\bar{N}_L$), are obtained after orthonormalization of the wave-function subspace (\ref{eq:Ansatz2}).

The total overlap with the exact states within the first excited, (quasi)degenerate manifold of the Hamiltonian (\ref{eq:H_EBH}) is 
\begin{equation}\label{eq:Overlap_Ex}
\bar{O}_{\bar{i},(\bar{t},\bar{U})}= \sqrt{\sum_{k=1}^{\bar{N}_L} |_{(\bar{t},\bar{U})}\langle\bar{\psi}_{Exact}^{(\bar{i})}|\bar{\psi}_{Ansatz}^{(k)}\rangle_{(\bar{t},\bar{U})}|^2},
\end{equation}
where $\bar{i}=1,...,\bar{N}_L$ denotes the states $|\bar{\psi}_{Exact}^{(\bar{i})}\rangle_{(\bar{t},\bar{U})}$ within the first excited states manifold.

\begin{figure}[b!]
 \includegraphics[width=\columnwidth]{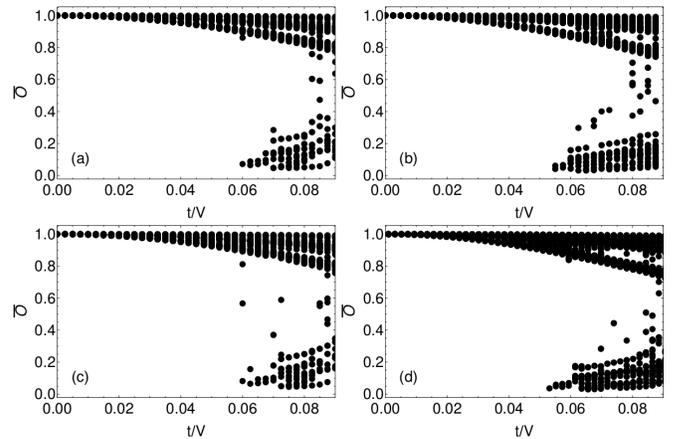}
\caption{\label{fig:E_Overlap} The ED results for the total overlaps (\ref{eq:Overlap_Ex}) of the $3L(L/2-1)$ exact, (quasi)degenerate states within the first excited states manifold of the Hamiltonian (\ref{eq:H_EBH}) at average filling of 
$\nu=3/2$ and with periodic boundary conditions, with the corresponding orthonormalized ansatz states. Here 
$t/V$ is the tunneling parameter with $V$ being the nearest-neighbor interaction, and the on-site interaction strength is $U/V=2$ ((a) and (b)) and $U/V=1.99$ ((c)and (d)). 
The system sizes are $L=10$ ((a) and (c)) and L=12 ((b) and (d)) lattice sites. 
  } 
 \end{figure}

The ED results for the overlaps (\ref{eq:Overlap_Ex}) are shown in Fig. \ref{fig:E_Overlap} for the system sizes $L=10$ and 12. For the values of the tunneling parameter $t/V<\bar{t}_c(L,\bar{U})$ the overlaps for all states within the first 
excited states manifold are $\simeq 1$. In other words, away from the degeneracy point at $U=2V$ and $t=0$, the nature and fractional charge of the domain walls do not change if $t/V<\bar{t}_c(L,\bar{U})$. This is of importance 
for actual experiments, where 
there is always some finite possibility for atoms tunneling between the lattice sites, and where the values of the on-site and nearest-neighbor interaction strengths can be 
tuned away from $U=2V$. 
 
 \begin{figure}[t!]
 \includegraphics[width=\columnwidth]{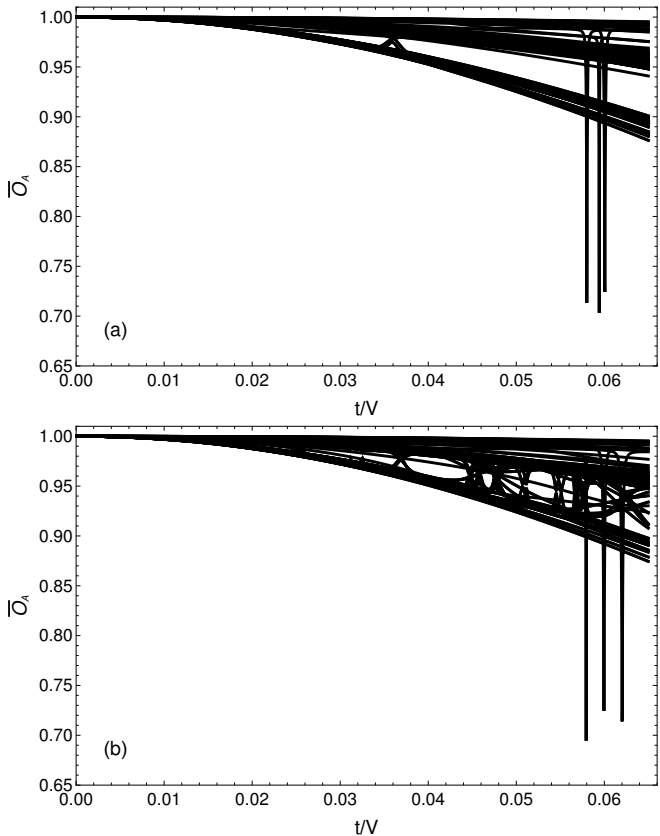}
\caption{\label{fig:Overlap_Adiabatic} The overlaps (\ref{eq:Overlap_Ex}) for $|\bar{\psi}_{Exact}^{(\bar{i})}\rangle_{\bar{t},\bar{U}}$ ($\bar{i}=1$,...,$\bar{N}_L$) taken to be 
the states adiabatically connected to the states within the first excited states manifold at $t=0$ (the states with one domain wall pair), and for the system size $L=10$ 
lattice sites with periodic boundary conditions. Here the on-site interaction strength is (a) $U/V=2$ and (b) $U/V=1.99$, with V being the nearest-neighbor interaction.  
  } 
 \end{figure}
 \begin{figure}[b!]
 \includegraphics[width=\columnwidth]{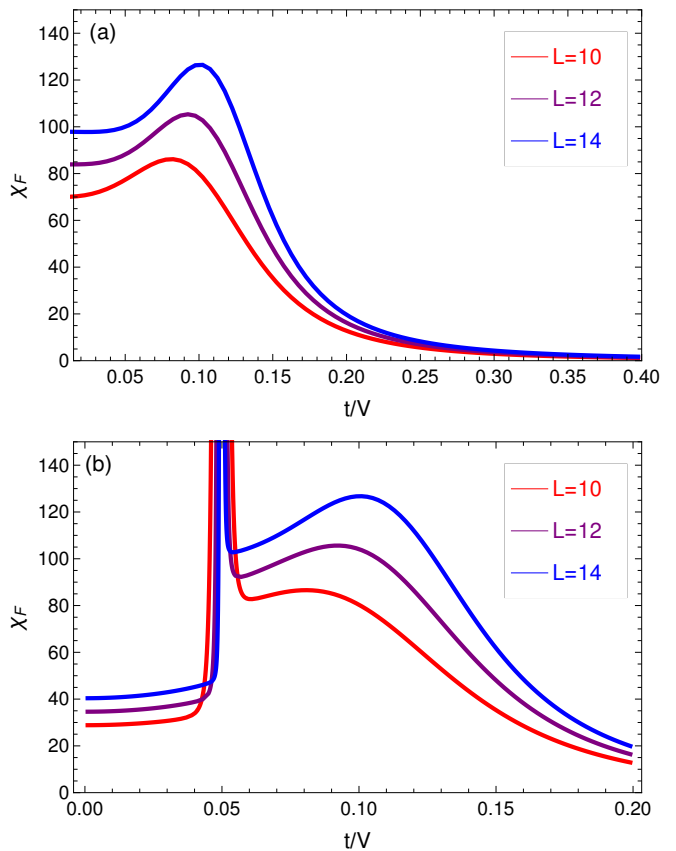}
\caption{\label{fig:Fidelity_Susceptibility} (Color online) The fidelity susceptibility $\chi_F$ (\ref{eq:Fidelity_Susceptibility}) as function of the tunneling parameter $t/V$, 
obtained by ED method for the system sizes $L=10$, 
12 and 14 lattice sites and with periodic boundary 
conditions. Here the average filling is $\nu=3/2$ atoms per lattice site and the on-site interaction strength is (a) $U/V=2$ and (b) $U/V=1.99$, with V being the 
nearest-neighbor interaction. 
} 
 \end{figure}
\begin{figure}[b!]
 \includegraphics[width=\columnwidth]{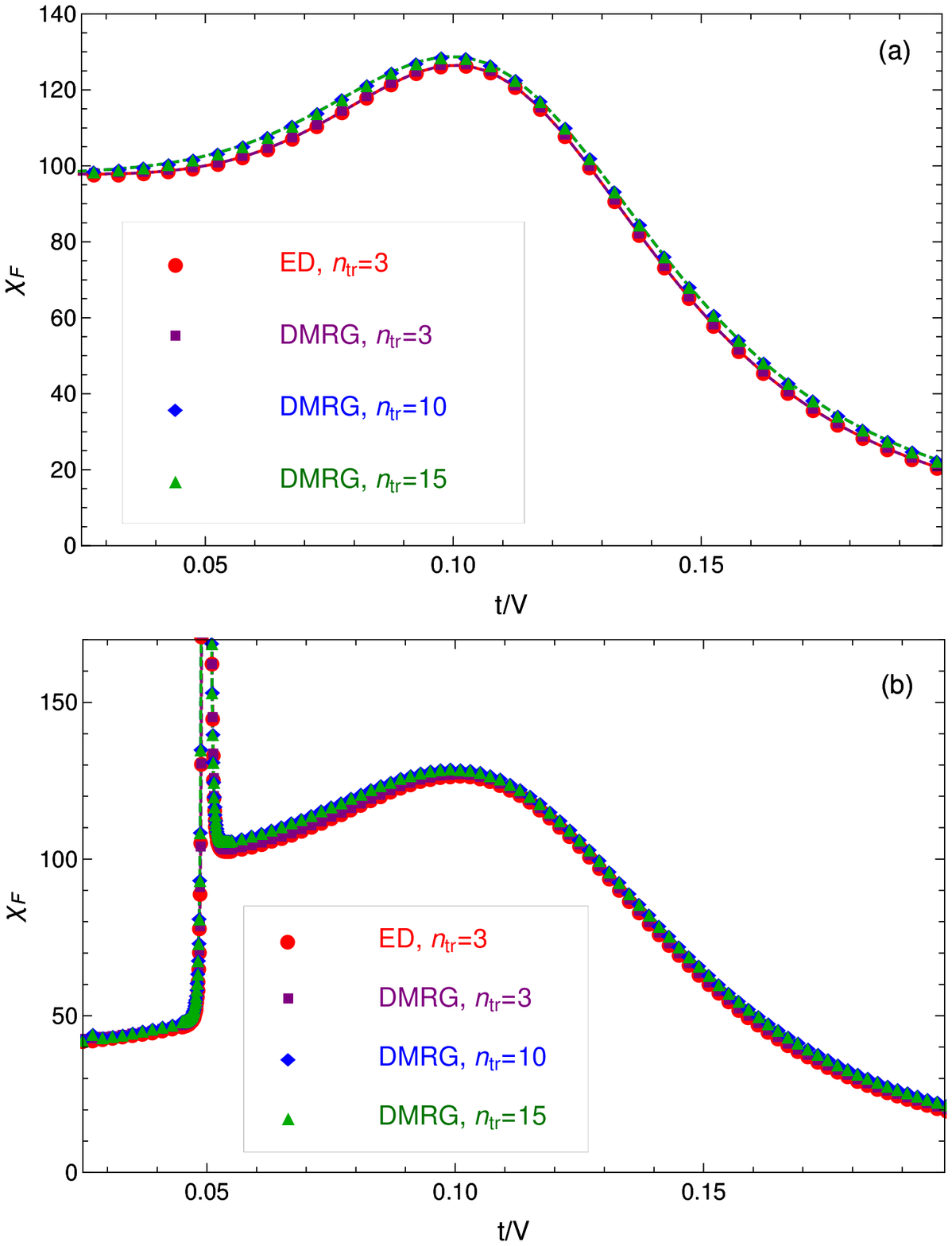}
\caption{\label{fig:Fidelity_SusceptibilityDMRG_L14} (Color online) The ED and DMRG results for the fidelity susceptibility $\chi_F$ 
(\ref{eq:Fidelity_Susceptibility}) as function of the tunneling parameter $t/V$, for the system size $L=14$ lattice sites and with periodic boundary conditions, the 
average filling $\nu=3/2$ atoms per lattice site, and with 
the local Hilbert space truncation to single site Fock states with at most $n_{tr}=3$, 10 and 15 atoms at each lattice site. Here the on-site interaction strength is 
(a) U/V=2 and (b) U/V=1.99, with $V$ being the nearest neighbor interaction. 
  } 
 \end{figure}

Sudden decrease of the overlap for some of the excited 
states at $t/V=\bar{t}_c(\bar{U},L)$ is related to the energy level crossings between the states within the first and second excited states manifolds (Fig. \ref{fig:EnergyLevels_L10_L12}). Namely, 
as pointed out in Ref. \cite{Wikberg}, moving away from the degeneracy point, where domain walls do not interact, introduces interaction between domain walls via a linear potential. The strength
and sign of the potential depends on the energy splitting between the CDW states that are degenerate at $U=2V$ and $t=0$. For $t/V>\bar{t}_c(L,\bar{U})$, some states with two $\pm q/2$ domain wall pairs are more energetically favorable 
than some of the states with one $\pm q/2$ domain wall pair due to an attractive linear potential between the domain walls which results in energy level crossings and sudden decrease of the overlap for some of the states within the first excited
states manifold.

We also note that the overlaps (\ref{eq:Overlap_Ex}) for $|\bar{\psi}_{Exact}^{(\bar{i})}\rangle_{\bar{t},\bar{U}}$ ($\bar{i}=1$,...,$\bar{N}_L$) taken to be 
the states adiabatically connected to the states within the first excited states manifold at $t=0$ (the states with one domain wall pair), also decrease significantly for some of these states when $t/V>\bar{t}_c(\bar{U},L)$, as shown in Fig. \ref{fig:Overlap_Adiabatic}.
In other words, the fractional domain walls do not have non-Abelian statistics for $t/V>\bar{t}_c(\bar{U},L)$, after the crossing between the states within different (quasi)degenerate manifolds.

 \begin{figure}[t!]
 \includegraphics[width=\columnwidth]{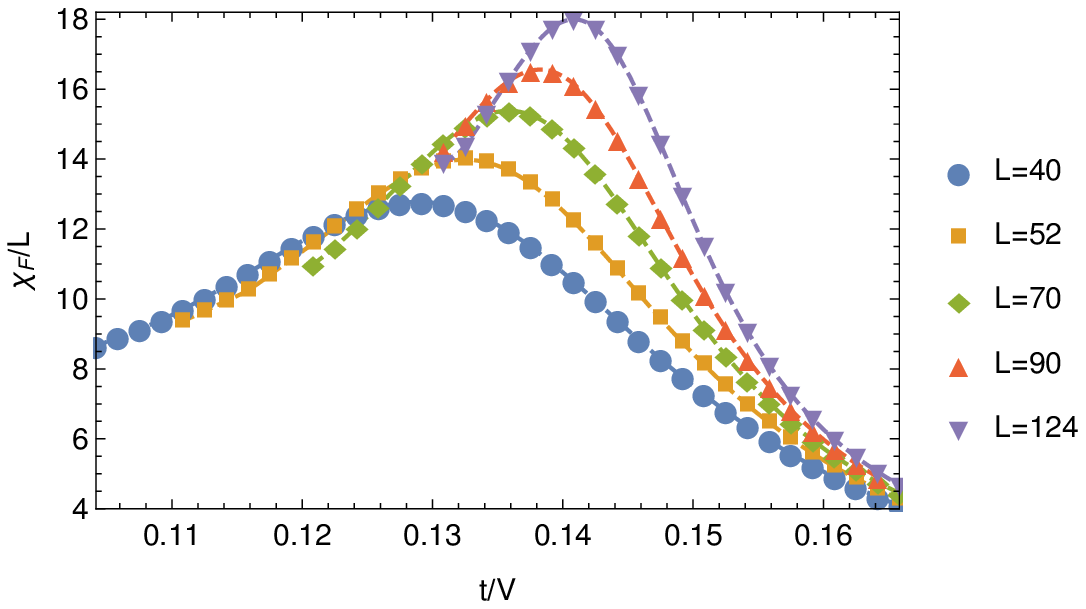}
\caption{\label{fig:Fidelity_SusceptibilityDMRG} (Color online) The DMRG results for the fidelity susceptibility $\chi_F$ 
(\ref{eq:Fidelity_Susceptibility}) as function of the tunneling parameter $t/V$, for the system sizes $L=40$ - $124$ lattice sites and with periodic boundary conditions, the 
average filling $\nu=3/2$ atoms per lattice site,
and with 
the local Hilbert space truncation to single site Fock states with at most $n_{tr}=4$ atoms at each lattice site. Here the on-site interaction strength is 
$U/V=2$ with $V$ being the nearest-neighbor interaction. 
  } 
 \end{figure}
 
In addition, for $U<2V$ increasing the tunneling strength $t/V$ induces the first order phase transition from $[30]$ ($[03]$) to $[21]$ ($[12]$) CDW state, 
as demonstrated previously using the Gutzwiller ansatz wave function \cite{Wikberg}. This first order transition, characterized by an energy level crossing, can be clearly seen in the fidelity 
metric \cite{Varney1,Duric2,Wang,Gu,Abasto,Eriksson,Varney2,Zanardi,Venuti,Rigol,Varney3}. If $|\psi_0(\bar{t})\rangle$ and 
$|\psi_0(\bar{t}+\delta\bar{t})\rangle$ are two ground states corresponding to slightly different values of the relevant parameter $\bar{t}=t/V$, the fidelity between these two 
ground states is defined as the modulus  of the overlap between the two states:
\begin{equation}\label{eq:Fidelity}
F(\bar{t},\bar{t}+\delta\bar{t})=|\langle\psi_0(\bar{t}+\delta\bar{t})|\psi_0(\bar{t})\rangle|.
\end{equation}
The fidelity (\ref{eq:Fidelity}) can further be rewritten as 
\begin{equation}\label{eq:Fidelity_susceptibility1}
F(\bar{t},\bar{t}+\delta\bar{t})=1-\frac{(\delta\bar{t})^2}{2}\chi_F(\bar{t})+...,
\end{equation}
where $\chi_F(\bar{t})$ is the fidelity susceptibility,
\begin{equation}\label{eq:Fidelity_Susceptibility}
\chi_F(\bar{t})=-\lim_{\delta\bar{t}\rightarrow0}\frac{2\ln F(\bar{t}+\delta\bar{t})}{(\delta\bar{t})^2}=-\frac{\partial^2 F(\bar{t}+\delta\bar{t})}{\partial(\delta\bar{t})^2}.
\end{equation}

The first order transition between two different CDW states is characterized by a singular peak in the fidelity susceptibility. Namely, since the overlap measures similarity 
between two states, it equals to one if two states are the same and zero if the states are orthogonal. Consequently, the fidelity shows a very sharp decrease at points where there is a level crossing between two 
orthogonal states, and decrease in the fidelity corresponds to a singular peak in the fidelity susceptibility. This singular peak can be clearly seen in Fig. \ref{fig:Fidelity_Susceptibility} 
at $t/V=\bar{t}_{CDW-CDW}(\bar{U},L)$ corresponding to the value of the tunneling parameter $t/V$ where there is an energy level crossing within the 
(quasi)degenerate ground-state manifold (Fig. \ref{fig:EnergyLevels_U=1.99V}).

Further increase of the value of the tunneling strength $t/V$ leads to a CDW to SF quantum phase transition of the BKT type, as it will be described in more details in the following section. This phase transition is characterized by a broader 
peak in the fidelity susceptibility which becomes sharper and sharper as the system size increases. This is clearly visible in figures \ref{fig:Fidelity_Susceptibility}, 
\ref{fig:Fidelity_SusceptibilityDMRG_L14} and \ref{fig:Fidelity_SusceptibilityDMRG}.

The transition is related to a level crossing between the states in the lowest energy, (quasi)degenerate manifold and the states within the first excited, 
(quasi)degenerate manifold at $t/V=\bar{t}_{CDW-SF}(\bar{U},L)$. The level crossings can be clearly seen in Fig. \ref{fig:EnergyLevels_U=2V} and Fig \ref{fig:EnergyLevels_U=1.99V} at values of $t/V$ which
coincide with the positions of the broader peaks in the fidelity susceptibility.

Our results thus demonstrate that the system undergoes a direct, BKT, CDW to SF quantum phase transition without intermediate SS phases between the CDW and SF 
regions of the phase diagram. This is in contrast with the results obtained previously within the Gutzwiller-ansatz wavefunction approach \cite{Wikberg}. Namely, previous 
results predicted two different SS phases, SS1 and SS2, separating CDW and SS regions of the phase diagram for $U=1.99V$. These SS phases are partially melted CDW phases, 
with SS1 and SS2 having different underlying CDW orders. The Gutzwiller-ansatz wave function calculations \cite{Wikberg} also predict CDW to SS1 and SS1 to SS2 transitions to 
be first order transitions, and SS2 to SF transition to be a second order transition. If SS phases were present in the phase diagram, these transitions would be clearly 
visible in the fidelity susceptibility. However, we do not find any signatures of such transitions and SS phases in our ED and DMRG results.

We also note that the Gutzwiller-ansatz wave function calculations were performed with the local Hilbert space truncation to single site Fock states $|n_i\rangle$ 
with at most $n_{tr}=30$ atoms at each lattice site ($0\leq n_i\leq n_{tr}$), while our ED and DMRG calculations were performed with $n_{tr}=3$. To check that increasing the 
truncation number $n_{tr}$ does not change qualitatively our results close to CDW to SF transition, we have performed additional calculations with $n_{tr}=10$ and 
$n_{tr}=15$. The results, shown in Fig. \ref{fig:Fidelity_SusceptibilityDMRG_L14} clearly demonstrate that increasing the truncation number $n_{tr}$ introduces only 
minor changes in the numerical values for the fidelity susceptibility 
and does not change our results qualitatively. We have also additionally verified that increasing the truncation number $n_{tr}$ to $n_{tr}\leq 10$ introduces only minor 
changes in our DMRG results for larger system sizes.

\section{Super-fluid to charge-density-wave quantum phase transition}
\label{sec:SF_CDW}
To further describe the SF to CDW quantum phase transition we calculate the density-density structure factor at wave number $k=\pi$
\begin{equation}\label{eq:structure_factor}
S_{\pi}=\frac{1}{N^2}\sum_{i,j=1}^Le^{i\pi(i-j)}\langle n_i n_j\rangle,
\end{equation}
the single particle correlation function 
\begin{equation}\label{eq:correlation_function}
\Gamma(|i-j|)=\langle a_i^{\dagger}a_j\rangle,
\end{equation}
and the associated system-size-dependent correlation length 
\begin{equation}\label{eq:correlation_length}
\xi_L=\sqrt{\frac{\sum_{i,j=1}^{L/2}(i-j)^2\langle a_i^\dagger a_j\rangle}{\sum_{i,j=1}^{L/2}\langle a_i^\dagger a_j\rangle}},
\end{equation}
for the system with L sites and N bosons and with periodic boundary conditions.

\begin{figure}[b!]
 \includegraphics[width=\columnwidth]{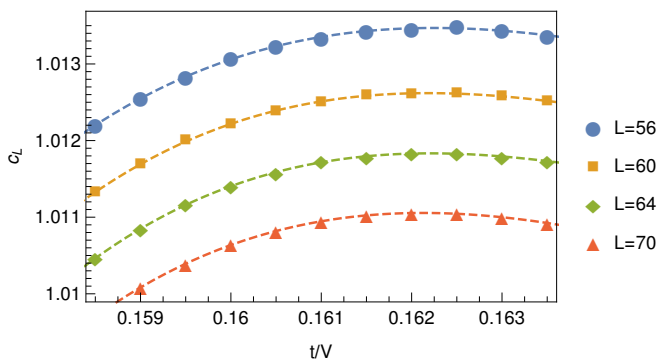}
\caption{\label{fig:CentralCharge} (Color online) The DMRG results for the central charge $c^{*}$ (\ref{eq:central_charge}) as a function of the tunneling 
parameter $t/V$ for several system sizes $L$ and with periodic boundary conditions. Here the on-site interaction strength is $U/V=2$ with $V$ being the nearest-neighbor interaction.
  } 
 \end{figure}
 \begin{figure}[t!]
 \includegraphics[width=\columnwidth]{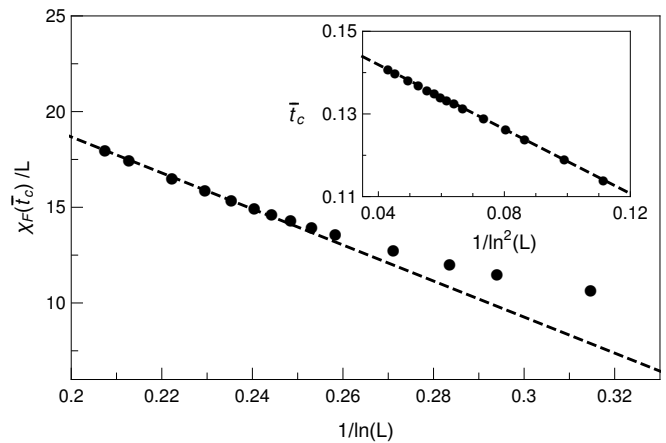}
\caption{\label{fig:Fidelity_Scaling} The finite-size scaling of the peak position $\bar{t}_c$ and amplitude $\chi_L(\bar{t}_c)$ of the fidelity susceptibility. The 
lines correspond to fits (\ref{eq:Fidelity_Scaling}) and (\ref{eq:Fidelity_tc}), where $A\approx$0.158, $B\approx$-0.39, $\chi_0\approx37.5$ and $\chi_1\approx$-94.2. 
 The data are for the system sizes $L=$20-124 lattice sites and with periodic boundary conditions. Here the on-site interaction strength is 
 $U/V=2$ with $V$ being the nearest-neighbor interaction. } 
 \end{figure}

We also calculate the von-Neumann block entanglement entropy
\begin{equation}\label{eq:entanglement_entropy}
S_L(l)=-\mbox{Tr}\left[\rho_l\ln\rho_l\right]
\end{equation}
where $\rho_{l}$ is the reduced density matrix for the block of length $l$. From 1+1 dimensional conformal field theory \cite{Calabrese,Ejima} it follows that the von Neumann 
entanglement entropy at a critical point has the form 
\begin{equation}\label{eq:entanglement_entropy2}
S_L(l)=\frac{c}{3}\ln\left[\frac{L}{\pi}\sin\left(\frac{\pi l}{L}\right)\right]+s_1
\end{equation}
for a system with periodic boundary conditions, with $s_1$ being a non universal constant and $c$ the central charge of the associated conformal field theory (CFT). 
Since DMRG calculations give the most precise data for $S_L(l)$ when $l=L/2$ \cite{Ejima, Nishimoto} , the most suited relation to determine the central charge is 
\begin{equation}\label{eq:central_charge}
c^* (L) \equiv \frac{3\left[S_L(L/2-1)-S_L(L/2)\right]}{\ln\left[\cos(\pi/L)\right]}, 
\end{equation}
where $c^*=c$ when the system is critical. The central charge provides definitive information about the universality class of a (1+1) - dimensional system \cite{Cardy}. 
Our results show that $c=1$ in the SF regime, where the low energy effective theory for the system, obtained by the Abelian bosonisation \cite{Giamarchi}, is the Tomonaga-Luttinger-liquid (TLL) 
Hamiltonian \cite{Haldane}. Within the non-Abelian bosonization \cite{Affleck2} the low energy theory of the SF phase is the Wess-Zumino-Witten (WZW) theory 
with topological coupling $k=1$ ($SU(2)_1$ WZW theory) \cite{Affleck} and the conformal anomaly parameter (central charge) $c=3k/(2+k)=1$ \cite{Affleck}.

The central charge can also be used to determine the critical point between TLL and gapped (or ordered) phases \cite{Nishimoto}. Namely the critical point corresponds to the 
maximum of $c^{*}$ (\ref{eq:central_charge}) as a function of $t/V$ \cite{Nishimoto}. The position of the maximum point, $(t/V)_c$, is independent of the system size for the model that 
we have considered (Fig. \ref{fig:CentralCharge}). Similar result was obtained for 1D half-filled spinless fermions with nearest-neighbor repulsion \cite{Nishimoto}. 

Our DMRG \cite{ITensor} results show that $(t/V)_c\approx 0.162$ (Fig. \ref{fig:CentralCharge}) for $U/V=2$. On the right-hand side of the maximum point $c^{*}$ approaches the value $c=1$ with increasing system size, 
and $c^{*}\rightarrow 0$ for the CDW gapped phase. In the DMRG calculations of the central charge dimensions of the matrices in the matrix product state (MPS) wave-function 
 were taken to be up to 2200 and $n_{tr}$=4.

To further characterize the nature of the SF to CDW quantum phase transition we consider the finite-size scaling of the fidelity susceptibility. Within the non-Abelian 
bosonisation approach it was shown that the fidelity susceptibility in the vicinity of a BKT transition has the following logarithmic finite-size scaling \cite{Sun} 
\begin{equation}\label{eq:Fidelity_Scaling}
\chi_L \simeq \chi_0 - \frac{\chi_1}{\ln(L/a)}+O\left[\frac{1}{\ln^2(L/a)}\right],
\end{equation}
where $a$ is the lattice cutoff. Also, the finite-size dependence of the peak position in the fidelity susceptibility, that signals 
the BKT transition, has the following form 
\begin{equation}\label{eq:Fidelity_tc}
\bar{t}_c\simeq A + B/\ln^2(L/a)+...,
\end{equation}
which can be obtained using scaling arguments on the gapped side of the BKT transition \cite{Sun}. Here $\bar{t}=t/V$. We fit our DMRG data for the fidelity susceptibility to these predicted 
finite size-scaling behaviors, and the results of these fits demonstrate good agreement with the theory (Fig. \ref{fig:Fidelity_Scaling}). This confirms that the SF to CDW
quantum phase transition is of the BKT type.

We also point out that $\bar{t}_c(L\rightarrow \infty) = A = 0.158 \pm 0.004$ which is consistent (within the error bars) with the 
value of $\bar{t}_c \approx 0.162$ obtained form the central charge. We have also studied the scaling of the energy gap in the vicinity of the transition \cite{Carrasquilla}.
The estimated transition point is then $\bar{t}_c(L\rightarrow\infty)=0.16\pm0.004$ in agreement with $\bar{t}_c$ obtained from the fidelity susceptibility studies.

We finally calculate the structure factor (\ref{eq:structure_factor}) close to the SF to CDW quantum phase transition to show that there is a direct phase transition from the 
SF to CDW phase. The non-zero structure factor characterizes the crystalline order, and in the case of direct transition from the SF phase has the form 
$S_{\pi}\sim \xi^{\gamma/\nu}\Phi(\xi/L)$ close to the transition \cite{Kuhner1,Kuhner2}, where $\Phi$ is a scaling function. For the case of a direct transition 
the structure factor is governed by the correlation length $\xi$ that characterizes SF order and diverges in the SF phase \cite{Kuhner1,Kuhner2}, which results in the mentioned form 
of the stucture factor close to the transition.

\begin{figure}[t!]
 \includegraphics[width=\columnwidth]{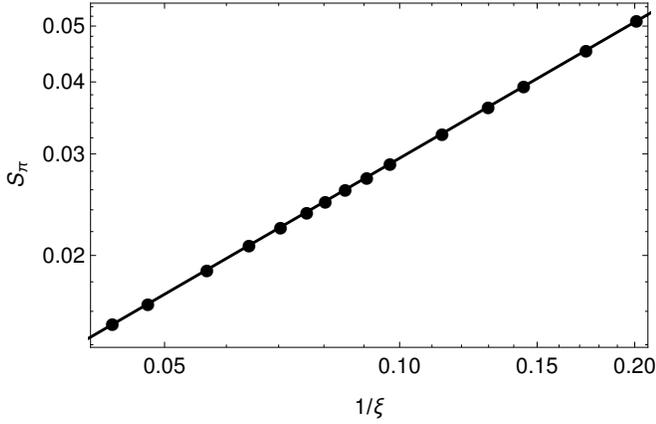}
\caption{\label{fig:Structure_Factor} The structure factor $S_{\pi}$ as a function $1/\xi$, where $\xi$ is the correlation length, at the BKT transition of the CDW phase 
($t/V\approx 0.158$). The slope is $\approx {-0.78}$ and $S_{\pi}\propto \xi^{-0.78}$. The data are for the system sizes $L=$20-124 lattice sites and with periodic 
boundary conditions. Here the on-site interaction strength is $U/V=2$ with $V$ being the nearest-neighbor interaction.
  } 
 \end{figure}

Also, the functional form of the structure factor cannot be transformed to a power law behaviour depending on $t/V$ since the 
correlation legth diverges like $\xi \propto \exp(const./\sqrt{(t/V)_c-(t/V)})$ at BKT type transition. Our results for the structure factor are shown in Fig. \ref{fig:Structure_Factor} 
and confirm that there is a direct SF to CDW transition without intermediate normal or supersolid phases. 
This is in agreement with previous results found by other authors \cite{Batrouni}.

\section{Protocol for braiding fractional domain walls} 
\label{sec:Braiding}
\begin{figure}[t!]
 \includegraphics[width=\columnwidth]{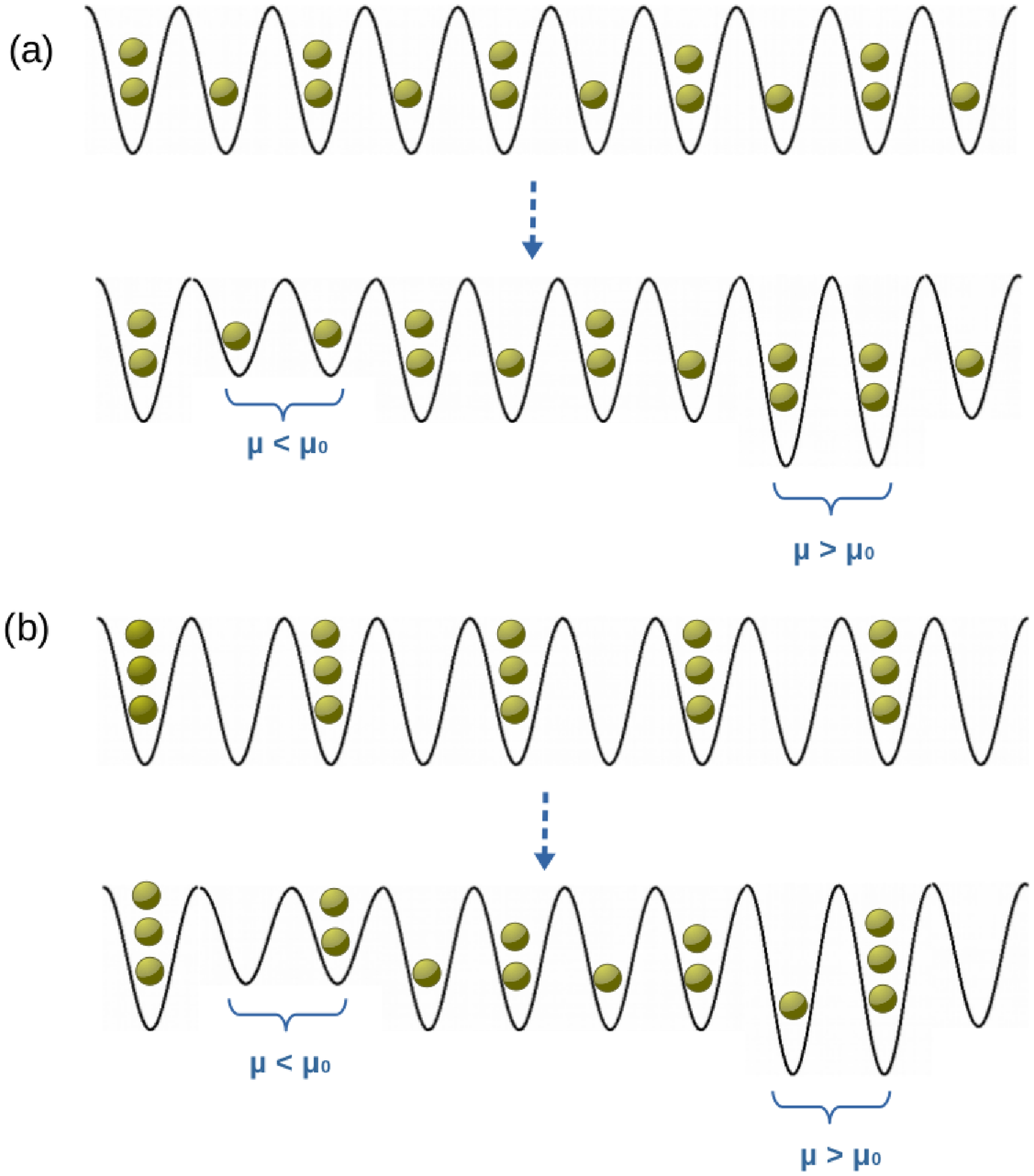}
\caption{\label{fig:FractionalDomainWalls} Schematic demonstration how local changes in the chemical potential can create robust SU(2)$_3$ Fibonacci anyon fractional domain walls 
which appear in a ground state configuration of the system, as suggested previously in Ref. \cite{Wikberg}. 
  } 
 \end{figure}

\begin{figure}[t!]
 \includegraphics[width=\columnwidth]{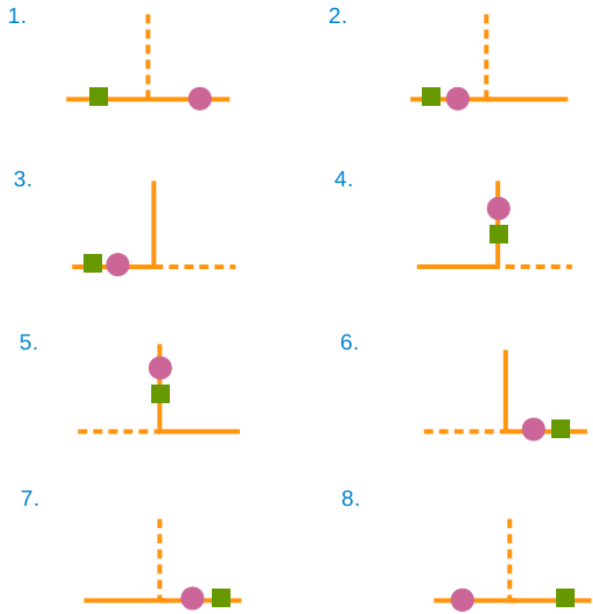}
\caption{\label{fig:Braiding} A T-junction which allows adiabatic exchange of two fractional domain walls. In each step of adiabatic exchange a dashed line represents a part
of the junction which is disconnected from the part of the junction represented by a solid line. Position of a domain wall on a 1D lattice represented by a solid line 
can be changed by an adiabatic change of the local chemical potential at corresponding sites of the initial and final positions of the domain wall 
(Fig. \ref{fig:FractionalDomainWalls}).
  } 
 \end{figure}
In order to use described fractional domain walls for quantum computation, that is to realize topological quantum gates, one needs to engineer states with robust fractional domain walls in a geometry where these domain 
walls can be interchanged in a controlled way (braided). To have robust fractional domain walls it is necessary to achieve that these domain walls appear in a ground state configuration of the system. For a 
fixed filling fraction this can be achieved by locally varying the chemical potential \cite{Wikberg} as illustrated in Fig. \ref{fig:FractionalDomainWalls}.

Namely, starting from the unperturbed initial configuration, 
increasing/decreasing the chemical potential on two neighboring sites creates $+q/2$/$-q/2$ fractional domain walls \cite{Wikberg}. The domain walls illustrated 
in Fig. \ref{fig:FractionalDomainWalls} are SU(2)$_3$ Fibonacci anyons similar to elementary excitations of the bosonic Read-Rezayi state \cite{Wikberg,ReadRezayi2,Cappelli}
\begin{eqnarray}\label{eq:RR}
&&\psi_{RR}=\mathcal{S} \left(\prod_{i_1<j_1}^{N/k}(z_{i_1}-z_{j_1})^2...\prod_{i_k<j_k}^{N/k}(z_{i_k}-z_{j_k})^2\right) \nonumber\\
&&\prod_{i<j}^N (z_i-z_j)^M e^{-(1/4)\sum_i|z_i|^2},
\end{eqnarray}
with $k=3$ and $M=0$ and where $\mathcal{S}$ denotes symmetrization over possible divisions of the atoms into $k$ clusters of the same size.

The adiabatic exchange (braiding) of the fractional domain walls is not possible in the strictly 1D system that we have considered. Therefore, to achieve controlled interchange 
of these non-Abelian defects, and realize topological quantum gates, several such 1D atomic quantum wires need to be combined into a 2D network where 1D wires are connected 
with T-junctions, as proposed previously for Majorana quantum wires \cite{Alicea}. A T-junction which allows adiabatic exchange of two fractional domain walls is illustrated 
in Fig. \ref{fig:Braiding}. A part of the T-junction which does not contain domain walls can be connected to or disconnected from the part of the junction with two domain walls 
by adiabatically switching on or off the tunneling between the neighboring sites of the two parts of the junction.

In Fig. \ref{fig:Braiding} a part of the junction that is 
disconnected from the rest of the junction in each step of the adiabatic exchange of two fractional domain walls is represented by a dashed line. A part of the junction which 
contains two domain walls is represented in each step by a solid line. Position of a domain wall on a 1D lattice represented by a solid line 
can be changed by an adiabatic change of the local chemical potential at corresponding sites of the initial and final positions of the domain wall (for example in the step 
from 1 to 2 in Fig. \ref{fig:Braiding}).

We also point out that braiding of fractional domain walls in a T-junction network requires only a few local operations on relevant sites where the local chemical potential 
and the tunneling strength between the two nearest-neighboring sites needs to be adiabatically changed in each step of the adiabatic exchange of these non-Abelian defects.

These adiabatic changes of the local chemical potential and the tunneling strength between the two nearest-neighboring sites can be achieved experimentally by using local 
site addressing tools available in current experiments with cold atoms and molecules \cite{Weitenberg,Fukuhara,Kraus}. In cold atom experiments these local operations can be 
realized in a controllable way by changing the intensity of tightly focused laser fields on the corresponding site or link \cite{Weitenberg,Fukuhara,Kraus}.  
\section{Conclusions}
\label{sec:Conculsions}
We have studied low energy properties of a system of dipolar lattice bosons trapped in a 1D optical lattice and at average filling $\nu=3/2$ atoms per lattice site. 
The system can be described by an extended Bose-Hubbard Hamiltonian 
with the on-site and nearest-neighbor interactions. Using ED and DMRG methods we have identified a region of the phase diagram where the system supports 
SU(2)$_3$ Fibonacci 
anyon excitations. The SU(2)$_3$ non-Abelian topological order of the exact wave functions of the Hamiltonian was demonstrated by calculating the overlaps with the 
ansatz wave functions which have SU(2)$_3$ topological order by construction.

Contrary to previous results obtained within the Gutzwiller ansatz wave-function approach \cite{Wikberg}, our ED and DMRG results demonstrated that for an average filling of 3/2 the system 
undergoes a direct, BKT, 
CDW to SF quantum phase transition when the tunneling strength between the nearest-neighboring sites of the lattice is increased above a certain critical value. We 
do not find any signatures of the SS phases in the phase diagram of the system, found in Ref. \cite{Wikberg} to appear between CDW and SF regions in the parameter space. However, the SS phases 
are predicted to appear at higher filling fractions \cite{Batrouni}.

We have also discussed a protocol which would allow creation of robust SU(2)$_3$ fractional domain walls in a ground state configuration of the system and their controlled 
adiabatic interchange (braiding), with potential application for fault tolerant, universal, topological quantum computation. The domain walls can be introduced in a ground 
state of the system by changing the local chemical potential on certain lattice sites \cite{Wikberg}, and braiding can be achieved by combining 1D atomic quantum wires into 
a 2D network where the 1D wires are connected with T-junctions, as previously proposed in the context of Majorana quantum wires \cite{Alicea}. Both creation and 
braiding of such domain walls are achievable with local site addressing tools available in current cold atom experiments \cite{Weitenberg,Fukuhara,Kraus}. 
\\
\begin{acknowledgments}
We thank Nicholas Chancellor, Omjyoti Dutta, Bogdan Damski, Benoit Gr\'emaud and Dominique Delande for very helpful suggestions and discussions. This  work  was  realized  under  National  Science  Centre 
(Poland)  Project  No.  DEC-2012/04/A/ST2/00088 and was supported in part by PL-Grid Infrastructure. We also acknowledge support from the EU Grant QUIC (H2020-FETPROACT-2014, Grant No. 641122).
\end{acknowledgments}

\end{document}